\documentstyle[sprocl]{article}

\def\Journal#1#2#3#4{{#1} {\bf #2}, #3 (#4)}
\def\PRD{{\em Phys. Rev.} D}
 
\begin{document}
\title{COLLIDING SPHERICALLY SYMMETRIC NULL DUST\\ IN EQUILIBRIUM}
 
\author{ L\'{A}SZL\'{O} \'{A}. GERGELY}
 
\address{Laboratoire de Physique Th\'{e}orique, Universit\'{e}
Louis Pasteur,\\ 3-5 rue de l'Universit\'{e} 67084 Strasbourg Cedex, France\\
and\\
KFKI Research Institute for Particle and Nuclear Physics,\\
Budapest 114, P.O.B 49, H-1525 Hungary}
 
\maketitle\abstracts{We present two recently obtained solutions of
the Einstein equations with spherical symmetry and one additional
Killing vector, describing colliding null dust streams.
}

\section{Introduction}
 
Null dust is a model for the incoherent superposition of waves
with random polarizations and phases, moving in a single
direction with the speed of light.
It also describes the high frequency (geometrical optics)
approximation for any type (including gravitational)
radiation. Null dust is characterized by the energy-momentum
tensor $T^{ab}=\rho l^a l^b$. Neither the energy density $\rho$
nor the null vector $l$ are uniquely determined and the former
can be absorbed in the latter by an appropriate rescaling.
Various exact solutions of the Einstein equations in the presence
of null dust are known \cite{Kramer1,Kramer2,KuBi}, but certainly the
most famous of them is Vaidya's shining star solution \cite{Vaidya}.
 
The energy-momentum tensor describing the collision of two streams
of null dust (with propagation vectors $k$ and $l$ pointing in
opposite spatial directions) is a sum of the two single null
dust energy-momentum terms:
\begin{equation}
T^{ab}=l^a l^b +k^a k^b \ . \label{T2}
\end{equation}
The Einstein equations for the matter source (\ref{T2}) in the
plane symmetric case were solved by Letelier \cite{Let1}. The
cylindrically symmetic case was studied by Letelier and
Wang \cite{Let2}. Progress in the numerical treatment of the spherically
symmetric case was achieved by Holvorcem, Letelier and
Wang \cite{Let3}. Under the assumption of staticity Date \cite{Date}
integrated half of the relevant equations. The solution was found
by Kramer \cite{KramerND1} and by us \cite{GergelyND1}, ours
containing two additional parameters. Later, assuming homogeneity
in addition to spherical symmetry, we have obtained a
Kantowski-Sachs type solution: a closed Universe filled
with colliding streams of radiation \cite{GergelyND2}.
The interpretation of such spherically symmetric solutions
as regions of space-time containing both escaping and
backscattered gravitational radiation was proposed by Poisson and
Israel \cite{PI}. Recently Kramer obtained static and stationary
solutions describing the collision of cylindrical null
dust beams \cite{KramerND2}.
 
Here we present briefly the spherically symmetric static
and homogeneous solutions.
 
\section{The spherically symmetric static and homogeneous
solutions}
 
The detailed derivation of these solutions was presented
elsewhere \cite{GergelyND1,GergelyND2}. The line element
including both the homogeneous case ($c=1$) and the static case
($c=-1$) can be given in the concise form:
\begin{equation}
ds^{2} =\frac{e^{L^{2}}}{cCR}
\left(dZ^{2}-2R^{2}dL^{2}\right)
+R^{2}d\Omega ^{2} \label{metricZL} \ .
\end{equation}
Here $R(L)$ is a transcendental function (depending on the error
function):
\begin{equation}
cCR =e^{L^{2}}-2L\Phi _{B}(L),\qquad
\Phi _{B}(L)=B+\int^{L}e^{x^{2}}dx>0\ ,  \label{RL}
\end{equation}
while $B$ and $C>0$ are constants of integration.
 
One may think of $Z$ as being a time coordinate $t$ in the
static case and a radial coordinate $r$ in the homogeneous
case. Conversely, the coordinate $L$ has the meaning of time
in the homogeneous case and is a radial coordinate in the
static case. However, for this interpretation to hold, a study of
the sign of the expression (\ref{RL}) of $cCR$ is necessary,
more precisely in both cases $R\ge 0$ should hold.
This requirement follows also from the energy
conditions \cite{GergelyND1,GergelyND2}.
 
First of all let us remark that there is a true singularity at
$R=0.$ This can be seen for example from the ill-behaveness at
$R=0$ of the Kretschmann scalar ${\cal R}_{abcd}{\cal R}^{abcd}
\propto R^{-6}$. Thus the domain of the coordinate
$L$ is bounded by its value(s) $L_0$ at $R=0$. The
constant $B$ is related to $L_0$ as:
\begin{equation}
B=\chi (L_0)=\frac{e^{L_0^{2}}}{2L_0}-\int^{L_0}e^{x^{2}}dx \ .
\label{BL}
\end{equation}
For each value of
the constant $B$ there are two values $L_{0\pm}$, a positive and a
negative one (Fig.1). These are situated symmetrically with
respect to the origin only for $B=0$.
 
Second, we express the function $cCR$ in the alternative form
\begin{equation}
cCR =2L\left[\chi (L)-\chi (L_0)\right] \ .\label{Rchi}
\end{equation}
Because $\chi (L)$ is monotonically decreasing:
$d\chi/dL=-{e^{L^2}}/{2L^2}<0$,
the domains of positivity of $R$ are
$L_{0-}<L<L_{0+}$ in the homogeneous case and $L<L_{0-}$ together
with  $L>L_{0+}$ in the static case.
 
\begin{figure}[tbh]
\hspace*{1in}
\special{hscale=25 vscale=25 hoffset=0.0 voffset=0.0
         angle=-90.0 psfile=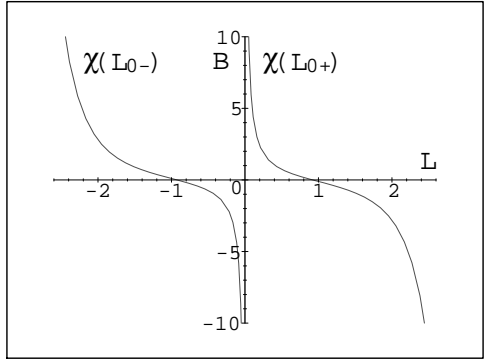}
\vspace*{2in}
\caption{The curves $B=\chi(L_{0\pm})$, corresponding to the
singularities $R=0$, devide the horizontal axis $L$ into domains
where $R>0,\ <0,\ >0$ (static case) or $R<0,\ >0,\ <0$
(homogeneous case).}
\end{figure}
 
\section{The interpretation of the solutions}
 
In the homogeneous case the time coordinate $L$ is bounded
both from above and below, the boundaries being singularities.
The radius function $R$ is increasing for $L\in (L_{0-},L^*)$, then
decreasing for $L\in (L^*,L_{0+})$, where $L^*$ is defined by
$\Phi_B(L^*)=0$.
The solution represents a closed Universe \cite{GergelyND2}.
From the symmetry group it is obvious that this is a
Kantowski-Sachs type solution, difficult to relate to
any physical situation.
 
We have analyzed in detail the static solution \cite{GergelyND1}
for $L>L_{0+}$.
The metric is not asymptotically flat and close to the axis $R=0$
(which is the singularity) its mass decreases to negative values.
Thus we run again into interpretational difficulties. However,
cutting the interior region along a timelike hypersurface, we can
glue the solution to some physically meaningful matter. We have
discussed the junction with a generic static interior \cite{GergelyND1}
by applying the Darmois-Israel \cite{Darmois,Israel} matching
procedure. In particular the junction algorithm with the interior
Schwarzschild solution \cite{Schw} has fixed the free parameters
$B$ and $C$ of the colliding null dust. The natural exteriors
are the incoming and outgoing Vaidya regions,
which were matched \cite{GergelyND1} along null hypersurfaces to
our solution, by applying the Barrab\`es-Israel junction
procedure \cite{BI}.
 
In the range $L<L_{0-}$ we have just an other copy of the static solution
with parameters $-B$ and $C$, and $-L$ as the radial variable.
Indeed, the metric (\ref{metricZL}) is invariant under the
interchange $(L,B)\to (-L,-B)$.
 
These solutions can be reinterpreted as anisotropic fluids with
radial pressures equal with their energy densities, and no
pressures in tangential directions to the spheres. An other approach we
already presented elsewhere \cite{GergelyND2} is based on
the connections with two-dimenssional dilatonic models. In this
context Mikovi\'c \cite{Miko} has given a solution in the form of
a perturbative series in powers of the outgoing part of the
energy-momentum tensor.

\section*{Acknowledgments}
 
This research has been supported by the OTKA grant D23744 and
by the Hungarian State E\"{o}tv\"{o}s Fellowship.


\section*{References}
\begin{thebibliography}{99}
 
\bibitem{Kramer1}  D. Kramer, H. Stephani, M.A.H. MacCallum, and
E. Herlt, {\em Exact Solutions of Einstein's Field Equations}
(Cambridge University Press, Cambridge, England, 1980).
 
\bibitem{Kramer2}  D. Kramer, in {\it Rotating Objects and
Relativistic Physics, Proceedings, El Escorial, Spain 1992}
, edited by J.J.Chinea and L.M. Gonz\'alez-Romero
(Springer Verlag, Berlin, 1993).
 
\bibitem{KuBi}  J. Bi\v {c}\'{a}k and K. Kucha\v {r},
    \Journal{\PRD}{56}{4878}{1997}.
 
\bibitem{Vaidya} P.C. Vaidya,
    \Journal{\em Proc.-Indian Acad. Sci., Sect. A}{33}{264}{1951}.
 
\bibitem{Let1} P.S. Letelier,
    \Journal{\PRD}{22}{807}{1980}.
 
\bibitem{Let2} P.S. Letelier and A. Wang,
    \Journal{\PRD}{49}{5105}{1994}.
 
\bibitem{Let3} P. R. Holvorcem, P.S. Letelier and A. Wang,
    \Journal{\em J. Math. Phys.}{36}{3663}{1995}.
 
\bibitem{Date} G. Date,
    \Journal{\it Gen. Relativ. Gravit.}{29}{953}{1997}.
 
\bibitem{KramerND1} D. Kramer,
    \Journal{\it Class. Quantum Grav.}{15}{L31}{1998}.
 
\bibitem{GergelyND1} L. \'A. Gergely,
    \Journal{\PRD}{58}{084030}{1998}.
The variables $R,r$ were denoted there by $r,r^*$.
 
\bibitem{GergelyND2} L. \'A. Gergely, submitted.
 
\bibitem{PI}  E. Poisson and W. Israel,
    \Journal{\PRD}{41}{1796}{1990}.
 
\bibitem{KramerND2} D. Kramer,
    \Journal{\it Class. Quantum Grav.}{15}{L73}{1998}.
 
\bibitem{Darmois} G. Darmois, in {\em M\'emorial des Sciences
Math\'ematiques} (Gauthier-Villars, Paris, 1927), Fascicule 25,
Chap. V.
 
\bibitem{Israel} W. Israel,
    \Journal{\em Nuovo Cimento B}{XLIV}{4349}{1966}.
 
\bibitem{Schw} K. Schwarzschild, {\em Sitz. Preuss. Akad. Wiss.},
424 (1916).
 
\bibitem{BI}  C. Barrab\`{e}s and W. Israel,
       \Journal{\PRD}{43}{1129}{1991}.
 
\bibitem{Miko} A. Mikovi\'c,
       \Journal{\PRD}{56}{6067}{1997}.
 
 
 
\end{thebibliography}
\end{document}